\documentclass[acmtog,authorversion,nonacm]{acmart}

\usepackage{balance}
\usepackage{dsfont}
\usepackage{color}
\usepackage{ifthen}
\usepackage{float}
\usepackage{alltt}
\usepackage{amsmath}
\usepackage{amsthm}
\usepackage{newlfont}
\usepackage{floatflt}
\usepackage{wrapfig}
\usepackage{fixltx2e}
\usepackage{subfig}
\usepackage{multirow}
\usepackage{booktabs}
\usepackage{cleveref}
\usepackage{algorithmic}
\usepackage[export]{adjustbox}
\usepackage{mathtools, cuted}
\usepackage{CJKutf8}
\usepackage[ruled]{algorithm2e}
\graphicspath{{figures/}}
\newtheorem{procedures}{Procedure}

\setcitestyle{square}

\SetAlFnt{\small}
\SetAlCapFnt{\small}
\SetAlCapNameFnt{\small}
\SetAlCapHSkip{0pt}
\IncMargin{-\parindent}

\let\oldcaption\caption
\renewcommand{\caption}[2][]{\oldcaption[#1]{{\em #1} #2}}

\definecolor{figred}{rgb}{1,0,0}
\definecolor{figgreen}{rgb}{0,0.6,0}
\definecolor{figblue}{rgb}{0,0,1}
\definecolor{figpink}{rgb}{1,0.63,0.63}

\newcommand{\pseudocode}{Algorithm}
\floatstyle{plain}
\newfloat{algorithm}{tbhp}{lop}
\floatname{algorithm}{\pseudocode}

\newcommand{\filename}[1]{\url{#1}}
\newcommand{\foldername}[1]{\url{#1}}

\let\oldparagraph\paragraph
\iffalse

\renewcommand{\paragraph}[1]{\oldparagraph{\textbf{#1}.}}
\else

\renewcommand{\paragraph}[1]{\oldparagraph{{#1}.}}
\fi

\ifdefined\email
\else
\newcommand{\email}[1]{\url{#1}}
\fi
\title[Toward Ubiquitous 3D Object Digitization: A Wearable Computing Framework for Non-Invasive Physical Property Acquisition]{Toward Ubiquitous 3D Object Digitization: A Wearable Computing Framework for Non-Invasive Physical Property Acquisition}

\author{Yunxiang Zhang}
\orcid{0000-0003-0189-1776}
\email{yunxiang.zhang@nyu.edu}
\affiliation{
 \institution{New York University}
 \country{USA}}

\author{Xin Sun}
\orcid{0000-0002-8710-2645}
\email{atlas.x.4@gmail.com}
\affiliation{
 \institution{Adobe Research}
 \country{USA}}

\author{Dengfeng Li}
\email{dengfli2-c@my.cityu.edu.hk}
\affiliation{
 \institution{City University of Hong Kong}
 \country{Hong Kong SAR, China}}

\author{Xinge Yu}
\orcid{0000-0003-0522-1171}
\email{xingeyu@cityu.edu.hk}
\affiliation{
 \institution{City University of Hong Kong}
 \country{Hong Kong SAR, China}}
 
\author{Qi Sun}
\orcid{0000-0002-3094-5844}
\email{qisun@nyu.edu}
\affiliation{
 \institution{New York University}
 \country{USA}}
\begin{teaserfigure}
\centering
\subfloat[non-invasive physical property acquisition]{
    \label{fig:teaser-left}
    \includegraphics[width=0.31\linewidth]{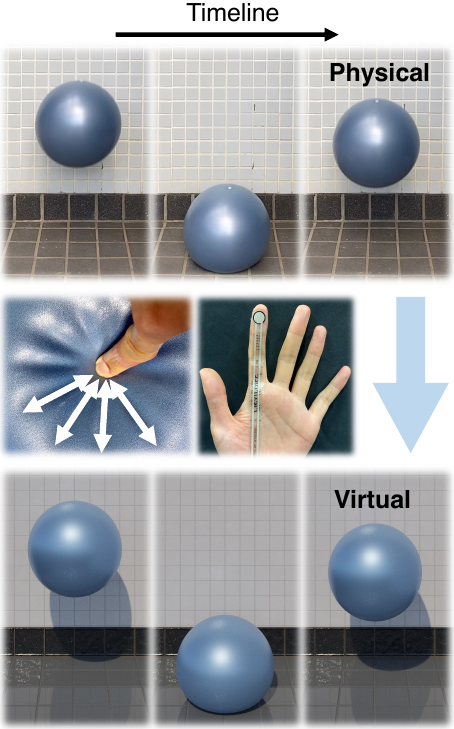}
  }
\subfloat[users interact with digitized objects exhibiting accurate physical properties in VR]{
    \label{fig:teaser-right}
    \includegraphics[width=0.65\linewidth]{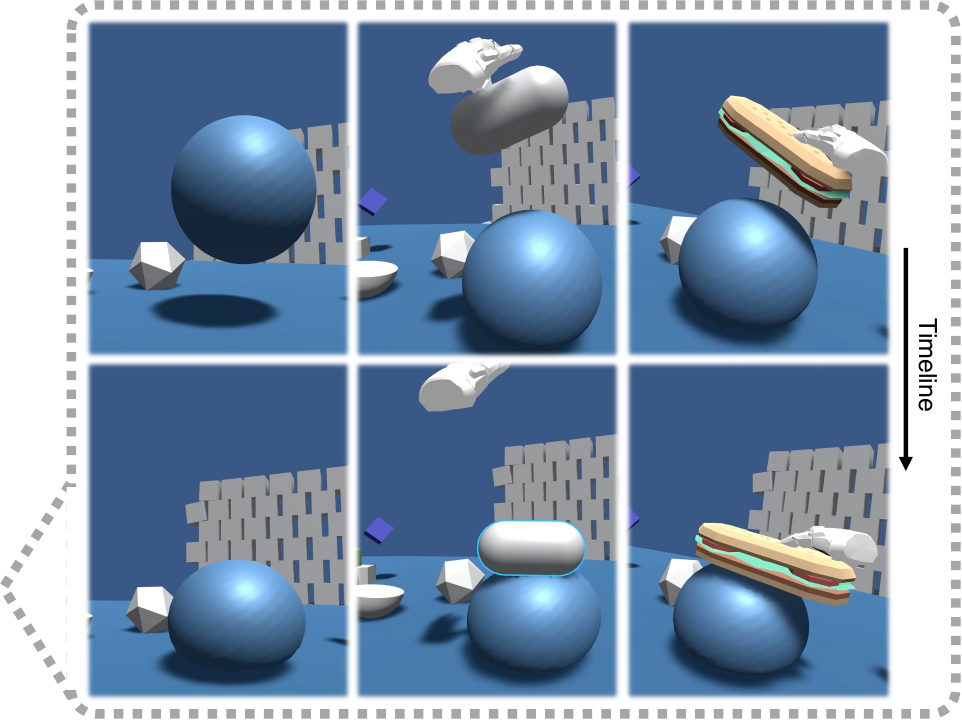}
  }
\caption{
Our wearable computing framework enables non-invasive digitization of deformable objects through a simple finger touch.
(a) By combining lightweight consumer-level electronics with computational physics models, we introduce a wearable and non-invasive computing framework that enables users to conveniently acquire the accurate physical properties (material elasticity and internal pressure) of daily soft objects.
(b) Under free-form user interventions, the resulting digital replicas exhibit consistent deformation behaviors as compared to their physical counterparts, unlocking realistic virtual interactions.
}
\label{fig:teaser}
\end{teaserfigure}
\begin{abstract}
Accurately digitizing physical objects is central to many applications, including virtual/augmented reality, industrial design, and e-commerce. Prior research has demonstrated efficient and faithful reconstruction of objects' geometric shapes and visual appearances, which suffice for digitally representing rigid objects. In comparison, physical properties, such as elasticity and pressure, are also indispensable to the behavioral fidelity of digitized deformable objects. However, existing approaches to acquiring these quantities either rely on invasive specimen collection or expensive/bulky laboratory setups, making them inapplicable to consumer-level usage.

To fill in this gap, we propose a wearable and non-invasive computing framework that allows users to conveniently estimate the material elasticity and internal pressure of deformable objects through finger touches. This is achieved by modeling their local surfaces as pressurized elastic shells and analytically deriving the two physical properties from finger-induced wrinkling patterns. Together with photogrammetry-reconstructed geometry and textures, the two estimated physical properties enable us to faithfully replicate the motion and deformation behaviors of several deformable objects. For the pressure estimation, our model achieves a relative error of $3.5\%$. In the interaction experiments, the virtual-physical deformation discrepancy measures less than $10.1\%$. Generalization to objects of irregular shape further demonstrates the potential of our approach in practical applications. We envision this work to provide insights for and motivate research toward democratizing the ubiquitous and pervasive digitization of our physical surroundings in daily, industrial, and scientific scenarios.
\end{abstract}

\begin{document}
\maketitle
\section{Introduction}
\label{sec:introduction}

Digitization is an essential technology underpinning various applications, such as preserving cultural heritage and artistic masterwork \cite{zabulis2022digitisation,stanco2017digital}, performing industrial assessments before mass production \cite{kritzinger2018digital,min2019machine}, running virtual laboratories for teaching and training \cite{kvedar2016digital,fogel2018artificial,davies2019digitization}, as well as enabling realistic interaction in virtual/augmented reality \cite{jiang2021douleur,speicher2017vrshop,khor2016augmented,bruno20103d}.
Creating realistic digital twins for physical objects requires quantitative knowledge of both their visual appearances (geometry, texture, etc) and physical properties (elasticity, pressure, etc) \cite{kapteyn2021probabilistic,kim2015physics}.
Current digitization systems primarily focus on the former \cite{geiger2011stereoscan,pollefeys2008detailed,kim20173d}. While this paradigm has been proven successful for digitizing rigid objects, it remains deficient in characterizing the behaviors of deformable ones. For instance, an inflated yoga ball with a wood texture may appear identical to a wooden ball of similar size. Their responses to compression or collision, however, are fundamentally disparate. The complex deformation behaviors of deformable objects demand additional information to replicate. Typical examples include the internal pressure of balls and the elastic modulus of clothing.
 
\begin{figure*}[t]
    \centering
    \includegraphics[width=0.99\linewidth]{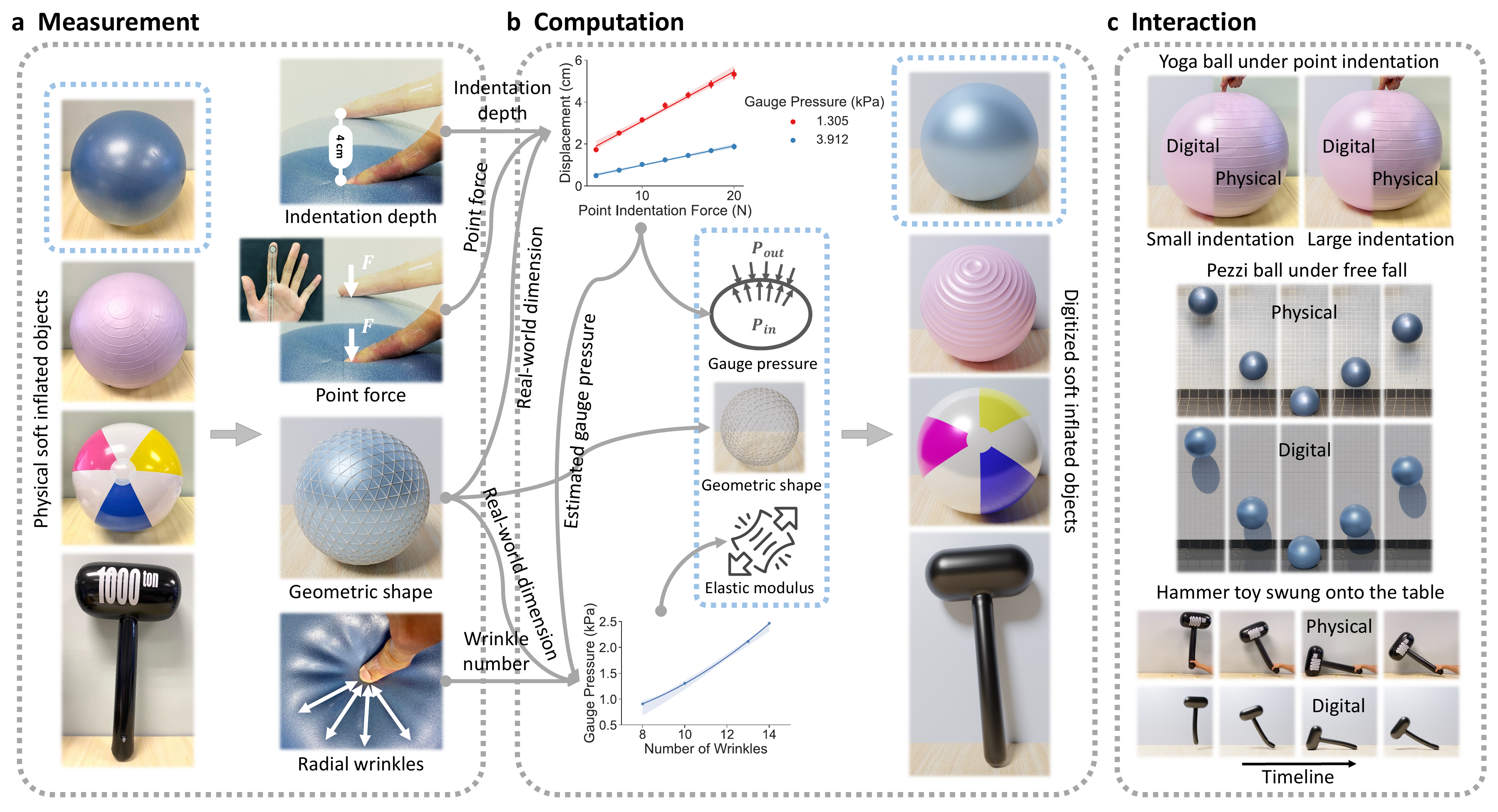}
    \caption{The overall workflow of our wearable and non-invasive computational digitization framework. (a) Given a target deformable object, we employ a photogrammetry application available on commercial mobile phones to reconstruct its geometry and textures, as well as a thin-film haptic sensor to measure the finger-exerted indentation force. (b) Using these non-invasively obtained quantities, we analytically compute the surface elastic modulus and internal pressure of the target object based on an inverse computational model. (c) We use all the measured and computed information above to create a faithful digital replica of the target object and interact with it in the virtual environment.}
    \label{fig:overview}
\end{figure*}

Unfortunately, existing approaches to obtaining these quantities, either through direct measurements, conventional model-based approximation \cite{becker2007robust,kauer2002inverse,ottensmeyer2001vivo,wang2015deformation}, or data-driven model fitting \cite{pai2001scanning,schoner2004measurement,bickel2009capture,bianchi2004simultaneous,feng2022learning}, inevitably involve high-cost measurement equipment, time-consuming computation procedures, or non-reversible operations that damage the target objects. For instance, the tensile test used for gauging a material's elastic modulus requires flat specimens cut from the target object \cite{kumar2014use}. Such destructive preparation is inapplicable in casual scenarios outside laboratory settings.
On an orthogonal research line, the computational physics community has extensively studied the deformation patterns that form on externally loaded shallow shells, such as wrinkling and buckling under point indentation \cite{cerda2002wrinkling,cerda2003geometry,li2011surface,vandeparre2011wrinkling}, and proposed analytical models that explain for the interplay between such patterns and the shell's physical properties \cite{vella2018regimes,box2019dynamics,jain2021compression}.

Inspired by the pressurized elastic shell model from computational physics \cite{vella2011wrinkling,vella2012indentation,taffetani2017regimes}, we propose a low-cost, lightweight, and efficient computational digitization framework that allows users to non-invasively estimate the surface elastic modulus and internal pressure of deformable objects through simple finger touches.
Specifically, by measuring the finger-exerted indentation force via wearable haptic sensors and observing the indentation-induced radial wrinkles, we establish an analytical inverse model to derive the two quantities of interest based on the indentation depth-force relationship and the wrinkle frequency. The proposed approach bypasses invasive tensile tests and high-cost manometers.
Together with photogrammetry-recovered geometry and textures \footnote{We employed a commercial 3D reconstruction application Polycam, which is based on photogrammetry, to recover the geometric shape and real-world dimension of target objects and stored the reconstruction results in the format of triangular meshes.}, the two estimated physical properties enable us to create faithful digital replicas for several daily objects. The workflow of our non-invasive digitization framework is illustrated in Figure~\ref{fig:overview}.
Comprehensive experimental results, both quantitative and qualitative, validate the efficiency and generality of our approach for preserving deformable objects' physical fidelity under external interventions, as reflected by their motion and deformation behaviors. We also demonstrate that users can experience natural and realistic interactions with the resulting digitized objects in VR (please refer to the supplementary video), such as ``bouncing-after-throwing'' and ``deformation-after-thumping''.
We envision our end-to-end framework to open up new possibilities for pervasively digitizing our physical surroundings while bypassing high-cost, time-consuming, or destructive laboratory-based measurements.

To summarize, our main contributions include:
\begin{itemize}
    \item an analytical inverse model for estimating the elastic modulus and internal pressure of deformable objects;
    \item a wearable and non-invasive computational digitization framework that requires minimal manual efforts and consumer-level hardware only;
    \item quantitative and qualitative evaluations using deformable objects of varying sizes and materials, as well as the demonstration of virtual interactions using the resulting digitized objects in VR.
\end{itemize}
\section{Related Work}

\subsection{Deformation Modeling}
\label{sec:prior-deformation}

Simulating realistic deformation of non-rigid materials requires careful choice of deformation models. Mass-spring models have been commonly exploited to represent and simulate objects of deformable nature thanks to their conceptual simplicity and computational efficiency~\cite{bridson2002robust,lloyd2007identification,delingette2008triangular,allard2006distributed}. Steinemann et al. used mass-spring models with distance-, surface-, and volume-preserving forces to characterize biological tissues for stable surgical simulation~\cite{steinemann2006hybrid}. Stanley and Okamura combined mass-spring simulation with haptic jamming to design a tangible and shape-changing human-computer interface~\cite{stanley2016deformable}. Leon et al. implemented a GPU-based mass-spring model to simulate the biomechanics of living tissues in real time~\cite{leon2010simulating}. Allard and Raffin incorporated mass-spring models into distributed settings for large-scale VR applications~\cite{allard2006distributed}. Delingette provided a formal connection between mass-spring models and continuum mechanics and produced isotropic deformations on unstructured meshes for nonlinear membrane modeling~\cite{delingette2008triangular}. Despite their flexibility, a major limitation of mass-spring models is the lack of intuitive connection between spring constants and materials' physical properties~\cite{nealen2006physically}.

A more principled yet sophisticated approach to deformable material modeling is to design constitutive models that accurately explain the various behaviors of deformable materials as reflected by empirical measurement data~\cite{xu2015nonlinear,arruda1993three,boyce2000constitutive,teran2005creating}. Each of these models has its specific set of free parameters that characterize the physical properties of the deformable material through their links to physical quantities such as Young's modulus, shear modulus, bulk modulus, and Poisson's ratio. In the field of interactive computer graphics, different constitutive models have been employed to capture the behaviors of hyper-elastic materials, including linear co-rotational model~\cite{muller2002stable}, Saint Venant–Kirchhoff model~\cite{barbivc2005real}, Neo-Hookean model~\cite{smith2018stable}, and Mooney–Rivlin model~\cite{wang2016descent}. For instance, researchers have leveraged these models to simulate tendons and muscles for hand animation~\cite{sueda2008musculotendon,zheng2022simulation}, soft tissue deformation for character animation~\cite{pai2018human,mcadams2011efficient}, volume-preserving flesh simulation~\cite{smith2018stable}, musculoskeletal simulation with heterogeneous materials~\cite{modi2021emu}, facial musculature simulation with passive tissue~\cite{sifakis2005automatic}. In this research, we adopted the Neo-Hookean model to perform triangular FEM simulations.

While simulating the general deformation behaviors of objects composed of elastic materials under external force loading with FEM is challenging, the computational physics community has researched several simplified cases of particular application value, including elastic thin films under tension/compression~\cite{cerda2002wrinkling,cerda2003geometry,song2008analytical,paulsen2016curvature} and elastic shells under point indentation \cite{vella2011wrinkling,vella2012indentation,taffetani2017regimes}. In particular, closed-form solutions have been proposed to explain the interplay between the deformation patterns that form on an externally loaded pressurized elastic shell and its physical properties \cite{vella2018regimes,box2019dynamics,jain2021compression}. Inspired by these models, we establish an analytical inverse model to computationally estimate the surface elastic modulus and internal pressure of deformable objects from non-invasively obtained measurements only.

\subsection{Elasticity Perception}

Besides suitable deformation models, carefully choosing their accompanying parameters is also crucial to the quality of the resulting simulation and requires lots of tuning efforts. The most straightforward way to bypass parameter tuning is to directly estimate them by measuring related physical quantities. For instance, tensile testing~\cite{davis2004tensile} gives Young's modulus and Poisson's ratio of an elastic material, which can, in turn, be used to compute the first and second Lam\'{e} parameters in Neo-Hookean models~\cite{ogden1997non}. However, measurement-based methods, such as tensile testing for gauging elastic modulus, often involve destructive and irreversible operations, which can be infeasible in many use cases.

Researchers have also explored data-driven approaches to fitting pre-defined deformation models. Pai et al. designed a scanning system that records the physical interaction behaviors of real deformable objects, including deformation response, contact textures, and contact sounds, for faithful reconstruction~\cite{pai2001scanning}. Lang et al. used Green's functions matrix representation to model elastic rods and achieved robust parameter estimation through customized regularization and fitting techniques~\cite{lang2002acquisition}. Becker and Teschner made use of linear FEM and quadratic programming to estimate the Young's modulus and Poisson's ratio of isotropic elastic materials~\cite{becker2007robust}. Kauer et al. modeled soft biological tissues as nonlinear viscoelastic continuums and combined axisymmetric FEM simulation with Levenberg–Marquardt algorithm to perform inverse parameter estimation~\cite{kauer2002inverse}. Kajberg and Lindkvist investigated the regime of large strains and presented a method for characterizing materials revealing plastic instability~\cite{kajberg2004characterisation}. Bickel et al. represented the deformation patterns observed on a real object as spatially varying stress-strain relationships for modeling and simulating nonlinear heterogeneous materials~\cite{bickel2009capture}. Frank et al. proposed to estimate the elasticity parameters of deformable objects by establishing the relationship between external force loading and resulting surface deformations. \cite{frank2010learning}. Similarly, Boonvisut and {\c{C}}avu{\c{s}}o{\u{g}}lu collected synchronized force loading and tissue deformation data using multi-axial force sensors and stereo cameras, and estimated the mechanical parameters of soft tissues via inverse FEM~\cite{boonvisut2012estimation}.
\section{Method}
\label{sec:method}

Commonly, deformable objects have complex geometric shapes and heterogeneous material compositions, making it hard to accurately model and analyze them. While the modeling of general deformable objects under arbitrary external loading remains a long-standing challenge, certain simplified cases of particular application value have been actively researched in computational physics, such as thin films under tension or compression~\cite{cerda2002wrinkling,cerda2003geometry,song2008analytical,paulsen2016curvature} and elastic shells under point indentation~\cite{vella2011wrinkling,vella2012indentation,taffetani2017regimes}. Therefore, instead of modeling a soft, inflated object holistically, we decompose the problem by focusing on its local deformation behaviors under finger-induced point indentation and infer the two physical properties of interest from there.

%%%%%%%%%%%%%%%%%%%%%%%%%%%%%%%%%%%%%%%%%%%%%%%%%%%%%%%%%%%%%%%%%%%%%

\subsection{Pressurized Elastic Shell Model}
\label{sec:theory}

Mathematically, we locate a convex surface region of relatively uniform curvature on the target object and model it as a spherical shallow shell of thickness $h$, curvature $1/R$, Young's modulus $E$ and Poisson's ratio $\nu$. Besides the inside/outside pressure difference, or gauge pressure, $P_{g}=P_{in}-P_{out}$, the point indentation also implies an external point force $F$ acting at the shell's apex (intersection between the shell and its axis of symmetry) in the normal direction. Shallow shell theory~\cite{timoshenko1959theory,calladine1989theory} provides a systematic way to analyze and understand this type of model, and the equations governing the shell's deformed geometry in the 2D polar coordinate system $(r,\theta)$ give:
\begin{equation}
\label{eq:shell-equation-1}
    B \nabla^{4} w + \frac{1}{R r} \frac{\mathrm{d}}{\mathrm{d} r}(r \psi) - \frac{1}{r} \frac{\mathrm{d}}{\mathrm{d} r}\left(\psi \frac{\mathrm{d} w}{\mathrm{d} r}\right) = P_{g} - \frac{\delta(r) F}{2 \pi r}
\end{equation}
\begin{equation}
\label{eq:shell-equation-2}
    \frac{1}{E h r} \frac{\mathrm{d}}{\mathrm{d} r}\left\{r \frac{\mathrm{d}}{\mathrm{d} r} \left[\frac{1}{r} \frac{\mathrm{d}}{\mathrm{d} r}(r \psi)\right]\right\} = \frac{1}{R} \nabla^{2} w - \frac{1}{2 r} \frac{\mathrm{d}}{\mathrm{d} r}\left(\frac{\mathrm{d} w}{\mathrm{d} r}\right)^{2}
\end{equation}
where $r$ denotes the radial length, $w(r)$ gives the vertical displacement, $\psi$ is the derivative of Airy stress with $\sigma_{\theta \theta} = \mathrm{d}\psi/\mathrm{d}r$ and $\sigma_{r r} = \psi / r$, and $B = E h^{3} / 12 (1 - \nu^{2})$ denotes the bending stiffness. The vertical displacement at the loading point, $w(0)$, is also commonly referred to as the indentation depth. The point force $F$ is incorporated into the equations through an indicator function $\delta(r) = \mathds{1}_{\{r=0\}}$. Note that Equations~\ref{eq:shell-equation-1} and~\ref{eq:shell-equation-2} explicitly relate the quantities of interest, the shell's elastic modulus $E, \nu$ and gauge pressure $P_{g}$, to its geometric features $h, R$ and deformation behaviors $w(r)$.

For the case of non-trivial gauge pressure that we consider, previous studies~\cite{vella2011wrinkling,vella2012indentation,jain2021compression,box2019dynamics} have introduced a characteristic radial length for shallow shells termed capillary length $l_{p}$ to non-dimensionalize Equations~\ref{eq:shell-equation-1} and~\ref{eq:shell-equation-2} into:
\begin{equation}
    \frac{1}{\tau^{2}} \nabla^{4} W + \frac{1}{\rho} \frac{\mathrm{d}}{\mathrm{d} \rho}(\rho \Psi) - \frac{1}{\rho} \frac{\mathrm{d}}{\mathrm{d} \rho}\left(\Psi \frac{\mathrm{d} W}{\mathrm{d} \rho}\right) = 1 - \frac{\delta(\rho) F}{2 \pi \rho P_{g}}
\label{eq:shell-equation-1-nd}
\end{equation}
\begin{equation}
    \frac{1}{\rho} \frac{\mathrm{d}}{\mathrm{d} \rho}\left\{\rho \frac{\mathrm{d}}{\mathrm{d} \rho} \left[\frac{1}{\rho} \frac{\mathrm{d}}{\mathrm{d} \rho}(\rho \Psi)\right]\right\} = \nabla^{2} W - \frac{1}{2 \rho} \frac{\mathrm{d}}{\mathrm{d} \rho}\left(\frac{\mathrm{d} W}{\mathrm{d} \rho}\right)^{2}
\label{eq:shell-equation-2-nd}
\end{equation}
where $l_{p} = (P_{g} R^{3} / E h)^{\frac{1}{2}}$, $\tau = P_{g} R^{2} / (E h B)^{\frac{1}{2}}$, $ \rho = r/l_{p}$, $W = w R/l_{p}^{2}$, and $\Psi = \psi/P_{g} R l_{p}$. Notably, $\tau^{-2}$, $W$, and $\rho$ are the dimensionless bending stiffness, vertical displacement, and radial length.

The above equations come with boundary conditions:
\begin{equation}
    \lim_{\rho \to 0} [\rho \Psi' - \nu \Psi] = 0, \quad \lim_{\rho \to \infty} W = 0, \quad \lim_{\rho \to \infty} \Psi = \frac{\rho}{2}
\label{eq:boundary-condition}
\end{equation}

Particularly, the 1st condition ensures zero horizontal displacements at the loading point, and the 3rd condition, which is the solution to Equations~\ref{eq:shell-equation-1-nd} and~\ref{eq:shell-equation-2-nd} in the absence of indentation ($F = 0$), enforces that the point force $F$ has no effect in regions that are far from the loading point.

In practice, most inflated objects have $E<1e8$ Pa and $h \ll R$. When inflated to the typical pressure range of their use cases, these objects largely fall into the regime $\tau \gg 1$ (e.g., the Pezzi ball that we use has $\tau \approx 40$ when inflated to $P_{g} \approx 1.3$ kPa), which means that the biharmonic term $\tau^{-2} \nabla^{4} W$ in Equation~\ref{eq:shell-equation-1-nd} can be safely ignored.

%%%%%%%%%%%%%%%%%%%%%%%%%%%%%%%%%%%%%%%%%%%%%%%%%%%%%%%%%%%%%%%%%%%%%

\subsection{Deformation Pattern: Radial Wrinkles}

As the point force $F$ increases, the indented shell will eventually surpass a critical state where radial wrinkles start to form around the loading point.~\cite{vella2011wrinkling} discovers by numerically solving Equations~\ref{eq:shell-equation-1-nd} and~\ref{eq:shell-equation-2-nd} under boundary conditions Equation~\ref{eq:boundary-condition} that this is because the hoop stress $\sigma_{\theta \theta} = \Psi'$ becomes compressive within an annular region $\rho \in [\rho_{\min}, \rho_{\max}]$ when the indentation depth $W(0) < -2.52$. Prior literature on the wrinkling of thin films has shown that the typical wavelength of wrinkles is characterized by a balance between bending and stretching~\cite{cerda2002wrinkling,cerda2003geometry}, which in this case can be expressed as $\lambda \sim (B R^{2} / E h)^{1/4}$. This result, combined with the fact that the radial extent of the wrinkled region is on the order of the capillary length $l_{p}$, gives $n \sim l_{p} / \lambda \sim \tau^{1/2}$. Numerical simulations further show that the scaling factor of this linear relationship is around 1.33~\cite{vella2011wrinkling}, and thus, we have:
\begin{equation}
    w^{c} \approx \frac{2.52 P_{g} R^{2}}{E h} \quad \Longrightarrow \quad E \approx \frac{2.52 P_{g} R^{2}}{h w^{c}}
\label{eq:modulus-estimation-1}
\end{equation}
\begin{align}
\begin{split}
    & n \approx 1.33 \tau^{\frac{1}{2}} = \frac{1.33 R}{h} \left(\frac{12(1-\nu^{2}) P_{g}^{2}}{E^{2}}\right)^{\frac{1}{4}} \\
    \Longrightarrow \quad & E \approx \sqrt{12(1-\nu^{2})} \left(\frac{1.33 R}{n h}\right)^{2} P_{g}
\label{eq:modulus-estimation-2} 
\end{split}
\end{align}
where $n$ denotes the number of radial wrinkles and $w^{c}$ denotes the critical indentation depth. Fig.~\ref{fig:elasticity}a shows the radial wrinkles that form on a beach ball under finger pressing. Note that more wrinkles emerge when we inflate the beach ball to a higher gauge pressure.

%%%%%%%%%%%%%%%%%%%%%%%%%%%%%%%%%%%%%%%%%%%%%%%%%%%%%%%%%%%%%%%%%%%%%

\subsection{Deformation Pattern: Inverted Spherical Cap}
\label{sec:cap}

In addition,~\cite{vella2012indentation} proves that the shape of the indented shell can be approximated by the inverted spherical cap~\cite{pogorelov1988bendings} when the dimensionless indentation depth $W(0) \ll -1$:
\begin{equation}
    W(\rho) = 
    \begin{cases}
    W(0) + \rho^{2}, & \rho \leq W(0)^{\frac{1}{2}} \\
    0, & \rho > W(0)^{\frac{1}{2}}
    \end{cases}
\end{equation}

As a result, the variation of the object's volume due to point indentation can be approximated by:
\begin{equation}
    \Delta V \sim \frac{\pi l_{p}^{4} W(0)^{2}}{2 R} = \frac{1}{2} \pi R w(0)^{2}
\label{eq:volume-variation}
\end{equation}

The work done by the point force $F$ in compressing the gas contained in the object is $P_{g} \Delta V \sim \frac{1}{2} \pi R w(0)^{2} P_{g}$. Differentiating Equation~\ref{eq:volume-variation} with respect to $w(0)$ gives:
\begin{equation}
    F \approx \pi k_{s} R P_{g} w(0) \quad \Longrightarrow \quad P_{g} \approx \frac{F}{\pi k_{s} R w(0)}
\label{eq:pressure-estimation}
\end{equation}
where $k_{s}$ denotes the linear scaling factor.

Under the regime of non-trivial point indentation (i.e., $W(0) \ll -1$), common soft inflated objects exhibit little discrepancy in $k_{s}$, as we observe in our experiments. Moreover, such a level of indentation can be easily approximated by finger pressing in practice.

%%%%%%%%%%%%%%%%%%%%%%%%%%%%%%%%%%%%%%%%%%%%%%%%%%%%%%%%%%%%%%%%%%%%%

\subsection{Computational Framework}
\label{sec:framework}

Assuming a universal $k_{s}$ among target objects, we devise a data-driven approach to estimating their gauge pressure $P_{g}$ by exploiting the linear relationship established in Equation~\ref{eq:pressure-estimation}. More precisely, using one or more pre-selected objects for data collection, we first train a linear regression model to estimate the constant $k_{s}$.
\begin{procedures}
\label{procedure1}
    Estimate the constant linear factor $k_{s}$.
    \begin{enumerate}
    \item Locate a convex surface region for point indentation and measure the local curvature $1/R$~\footnote{We compute the local curvature of the surface region of interest using the object's mesh recovered by photogrammetry-based 3D reconstruction application Polycam.};
    \item Measure the ground-truth gauge pressure $P_{g}$;
    \item Gradually indent the surface and collect a sequence of $\{F, w(0)\}_{i=1}^{N_{1}}$ pairs during the process;
    \item Perform linear regression on $\{F, w(0)\}_{i=1}^{N_{1}}$ to obtain an estimate of $k_{s} P_{g} \approx F/\pi R w(0)$, denoted by $\hat{P_{g}}$;
    \item Repeat from steps 2) to 4) using different gauge pressure to collect a sequence of $\{P_{g}, \hat{P_{g}}\}_{j=1}^{N_{2}}$ pairs;
    \item Perform linear regression on $\{P_{g}, \hat{P_{g}}\}_{j=1}^{N_{2}}$ to obtain an estimate of $k_{s} \approx \hat{P_{g}}/P_{g}$.
\end{enumerate}
\end{procedures}
Note that this process of estimating $k_{s}$ only needs to be performed once at the preparation stage and can be considered as a calibration step for the linear model in Equation~\ref{eq:pressure-estimation}. Given a previously unseen object with unknown gauge pressure at the deployment stage, we can now non-invasively estimate its gauge pressure $P_{g}$.
\begin{procedures}
\label{procedure2}
    Estimate the gauge pressure $P_{g}$.
    \begin{enumerate}
    \item Locate a convex surface region for point indentation and measure the local curvature $1/R$;
    \item Gradually indent the surface and collect a sequence of $\{F, w(0)\}_{i=1}^{N_{1}}$ pairs during the process;
    \item Perform linear regression on $\{F, w(0)\}_{i=1}^{N_{1}}$ to obtain an estimate of $k_{s} P_{g} \approx F/\pi R w(0)$, denoted by $\hat{P_{g}}$;
    \item Use the previously estimated universal $k_{s}$ to compute the gauge pressure $P_{g} = \hat{P_{g}}/k_{s}$.
\end{enumerate}
\end{procedures}

Besides the gauge pressure $P_{g}$, local curvature $1/R$, and indentation depth $w(0)$, we can measure the thickness $h$ with a vernier caliper by squeezing and flattening a small part of the object's surface. It then remains to estimate the elastic moduli $E$ and $\nu$. While Equations~\ref{eq:modulus-estimation-1} and~\ref{eq:modulus-estimation-2} are independent with two unknowns, we find the onset of wrinkling ambiguous to determine in practice, which makes the measurement of critical indentation depth $w^{c}$ unreliable. Furthermore, given the inversely proportional relationship between $E$ and $w^{c}$, the measurement error in $w^{c}$ can directly leak to the estimation of $E$ and lead to relative error on the order of $100\%$. Therefore, Equation~\ref{eq:modulus-estimation-1} shall not be applied to solve for $E$ and $\nu$. In contrast to the instability of $w^{c}$, we find the number of wrinkles $n$ (beyond the onset of wrinkling) to be consistent across different trials and easy to determine. With two unknowns $E$, $\nu$ and one useful equation, we choose to reduce the problem dimension by assuming a typical value of $\nu$ for elastic materials ($\nu = 0.4$ throughout our experiments) and use Equation~\ref{eq:modulus-estimation-2} to estimate $E$ only. This reduction is valid because, in practice, the elastic materials used for manufacturing the surfaces of soft inflated objects differ significantly less in $\nu$ (around 0.3 -- 0.4) than in $E$ (around $1e5$ -- $1e8$ Pa). In addition, if we assume $\nu$ as known and use Equation~\ref{eq:modulus-estimation-2} to compute $E$, then altering the value of $\nu$ from 0.3 to 0.4 only changes $E$ by $4.1\%$. Note that $\nu$ has much less of an impact on the deformation of elastic shells than $E$.
\begin{procedures}
\label{procedure3}
    Estimate the elastic modulus $E$.
    \begin{enumerate}
    \item Locate a convex surface region for point indentation and measure the local curvature $1/R$;
    \item Estimate the gauge pressure $P_{g}$ via Procedure~\ref{procedure2};
    \item Indent the surface until the radial wrinkles that form around the loading point stabilize and record the number of wrinkles; 
    \item Compute the elastic modulus $E$ via Equation~\ref{eq:modulus-estimation-2}.
\end{enumerate}
\end{procedures}
\section{Evaluation}
\label{sec:evaluation}

\begin{figure*}[t]
    \centering
    \includegraphics[width=0.99\linewidth]{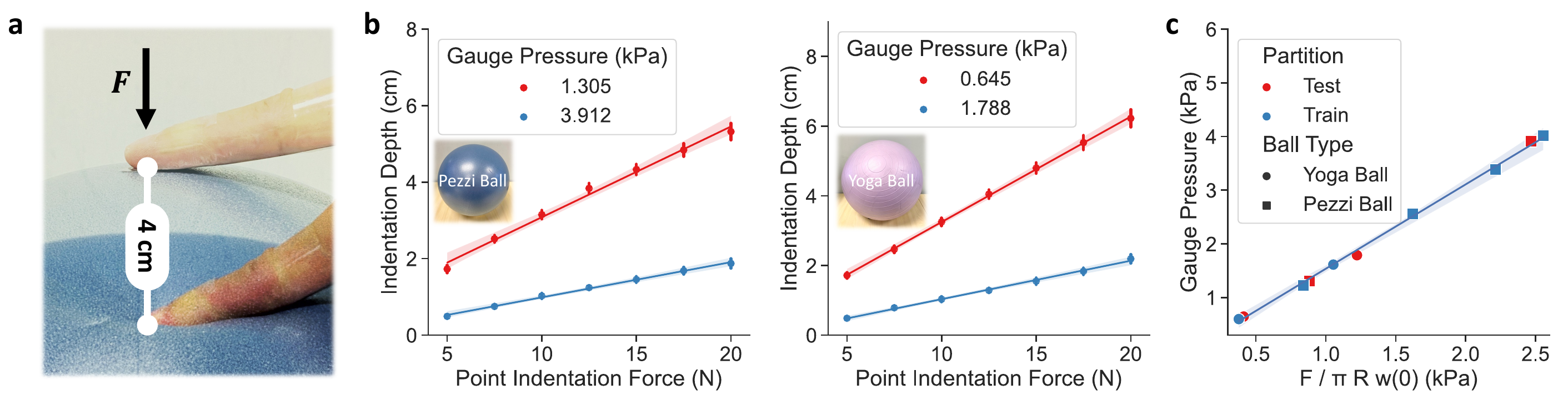}
    \caption{Estimation of gauge pressure via finger-induced point indentation. (a) Finger-induced point indentation. (b) Linear regression between point force $F$ and indentation depth $w(0)$ for estimating $\hat{P_{g}}$. Each point denotes the mean indentation depth of three trials sharing the same force intensity and the vertical bar shows the standard deviation. The translucent band around a regression line gives the $99\%$ confidence interval for the estimate. (c) Linear regression between $\hat{P_{g}} \approx F/\pi R w(0)$ and $P_{g}$ for estimating $k_{s}$. `Train' denotes the data used for regressing $k_{s}$; `Test' denotes the data used for evaluation.}
    \label{fig:pressure}
\end{figure*}

In the following, we first present the quantitative results of non-invasively estimating the gauge pressures and surface elastic moduli of several balls taking varying sizes and materials under the computational framework described in Section~\ref{sec:framework}. After that, we combine the two estimated physical properties with photogrammetry-reconstructed geometry to digitally replicate a Pezzi ball and qualitatively study the virtual-physical consistency through bouncing experiments. Finally, we apply our method to digitize an inflated hammer toy to demonstrate the generality of our approach.

%%%%%%%%%%%%%%%%%%%%%%%%%%%%%%%%%%%%%%%%%%%%%%%%%%

\subsection{Estimating Gauge Pressure}
\label{sec:pressure}

We select a Pezzi ball and a yoga ball with removable plugs to flexibly vary their gauge pressures using an air pump while measuring the corresponding ground-truth values using a digital manometer for evaluation. To practically obtain the point force $F$, we adopt piezoresistor-based haptic sensors shaped into flexible thin films and capable of providing sensitive responses to stress loading. With the haptic sensors mounted on the curved surfaces of fingertips, we can readily approximate point indentation by finger poking as illustrated in Figure~\ref{fig:pressure}a.

To reduce the variance in estimated $\hat{P_{g}}$ due to measurement errors in $F$ and $w(0)$ for both step 4) in Procedure~\ref{procedure1} and step 2) in Procedure~\ref{procedure2}, we collect the indentation depth $w(0)$ corresponding to a sequence of evenly-spaced point force $F$ (range from 5N to 20N, with a spacing of 2.5N). As can be observed in Figure~\ref{fig:pressure}b, the linear tendency between $F$ and $w(0)$ is clear for both balls under varying gauge pressure, and the relative regression error is no larger than 10\% on average. Repeating the above process six times using the Pezzi ball and the yoga ball under different gauge pressure, we reach a sequence of $\hat{P_{g}}, P_{g}$ and obtain a robust estimate of the linear factor $k_{s}=0.64$, i.e., step 6) in Procedure~\ref{procedure1}. Using the estimated $k_{s}$, which is approximately constant when $W(0) \ll -1$, we can now vary the gauge pressure of either ball to an unknown value and perform the estimation as detailed in Procedure~\ref{procedure2}. As shown in Figure~\ref{fig:pressure}c, we achieve a relative error of $3.51\% \pm 2.23\%$ on the test points. Note that our method neither relies on the manometer nor the fact that the object has removable plugs at the deployment stage, thus achieving non-invasive estimations of gauge pressure. Many real-world inflated objects are equipped with anti-leakage valves that prevent the use of a manometer, in which case our non-invasive method is most necessary and beneficial. In the following, we demonstrate that it indeed generalizes well to those cases.

%%%%%%%%%%%%%%%%%%%%%%%%%%%%%%%%%%%%%%%%%%%%%%%%%%

\begin{figure*}[t]
    \centering
    \includegraphics[width=0.99\linewidth]{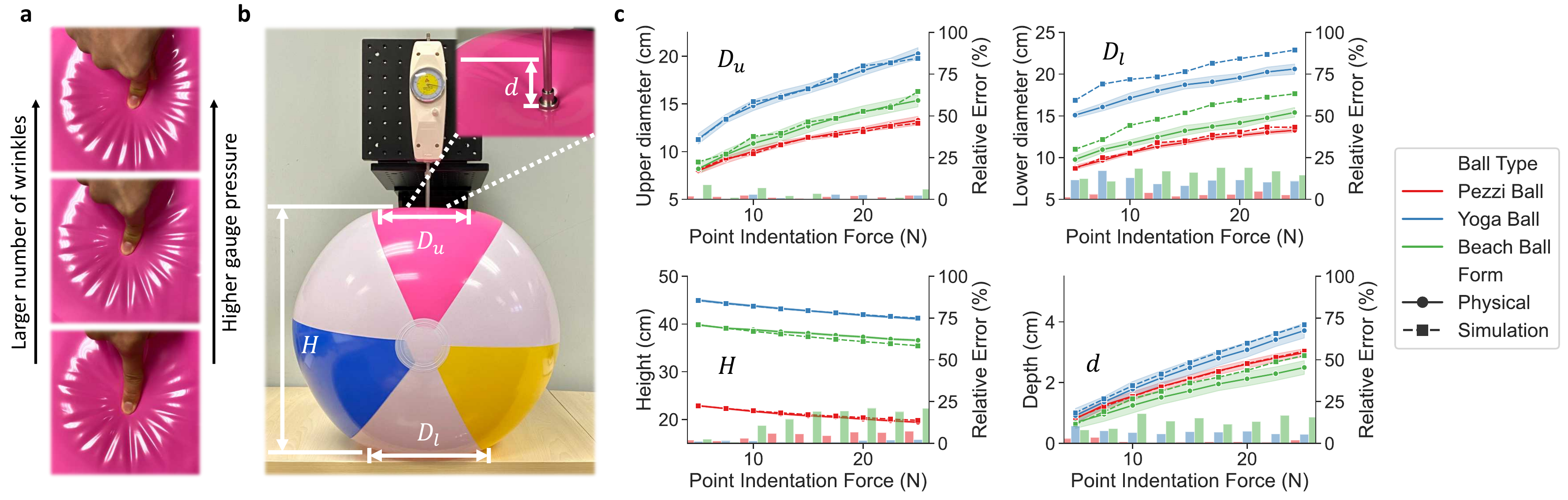}
    \caption{Evaluation of estimated elastic modulus $E$. (a) Formation of radial wrinkles on a beach ball being poked by fingers. Higher gauge pressure leads to more wrinkles. (b) Experimental setup for controlled vertical point indentation and illustration of the quantities being compared for evaluation. (c) Comparison between real balls deformed by vertical point indentation and their digitized counterparts. The left y-axis of each sub-figure measures the lengths shown by the line plots, while the right y-axis measures the relative errors shown by the bar plots. Each point denotes the mean of three trials sharing the same force intensity, and the translucent band around a line plot shows the standard deviation.}
    \label{fig:elasticity}
\end{figure*}

\begin{figure*}[t]
    \centering
    \includegraphics[width=0.99\linewidth]{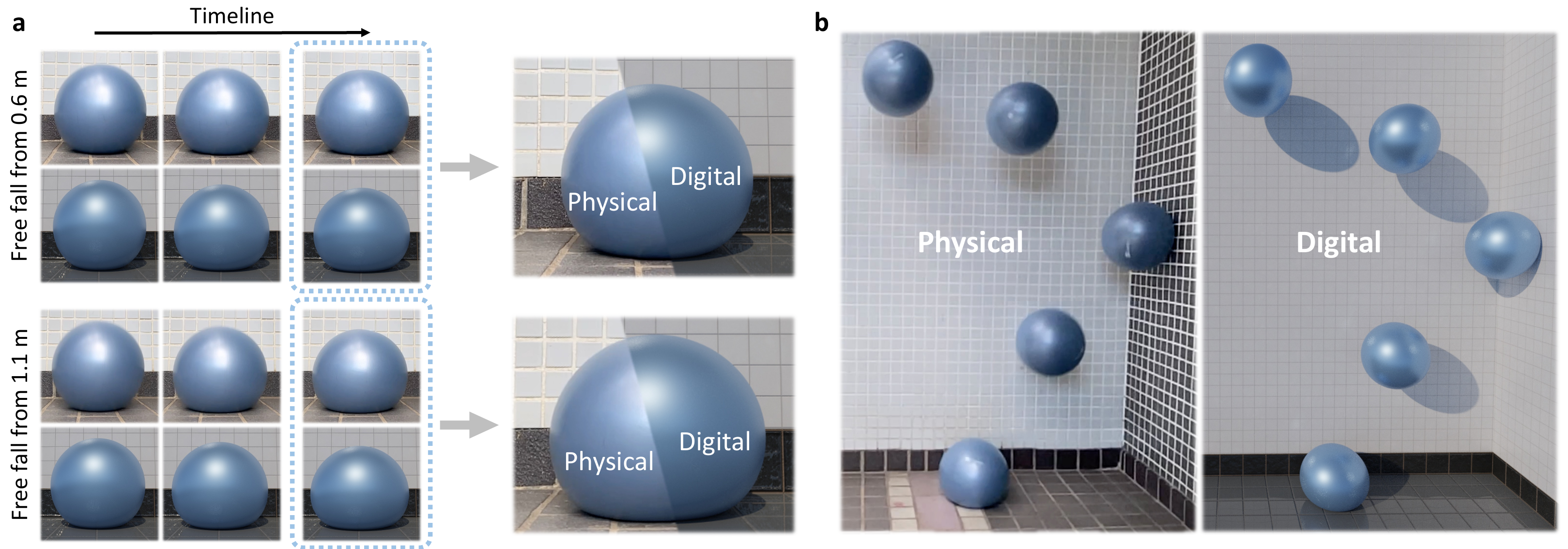}
    \caption{Dynamic behaviors of a digitized Pezzi ball. (a) Comparison between the maximally deformed shape of a real Pezzi ball undergoing a free fall and its digital counterpart. (b) Comparison between the same Pezzi ball bouncing against the wall and the ground in sequence and its digital counterpart.}
    \label{fig:bouncing}
\end{figure*}

\subsection{Estimating Elastic Modulus}
\label{sec:modulus}

While we can use manometers to determine the gauge pressures of the Pezzi ball and yoga ball for evaluation, this approach does not apply to general inflated objects with anti-leakage design. Moreover, directly measuring the elastic modulus $E$ via the tensile test will cause irreversible damage to the objects. Instead of establishing direct evaluations by comparing estimated values against the ground truth, we study to what extent do estimated $P_{g}$ and $E$, given by Procedures~\ref{procedure2} and~\ref{procedure3}, enable us to create realistic digital copies of real objects. More precisely, we first perform controlled physical interactions with real objects, then apply identical interactions to their digital replicas, which are controlled by a physics-based simulation algorithm fed with estimated $P_{g}$ and $E$. The simulation algorithm that we implement is a nonlinear finite element method (FEM) with triangle elements~\cite{zienkiewicz1977finite} (as explained in Section \ref{sec:framework}, we set the Poisson's ratio $\nu$ to be 0.4 for all deformable objects in our experiments). The geometric deformations induced by the interactions are used as a proxy to quantitatively evaluate the virtual-physical consistency as determined by the gauge pressure $P_{g}$ and elastic modulus $E$. In fact, creating physical realism for virtual objects under all sorts of interactions is also the ultimate goal of digitization for scenarios such as VR/AR.

To ensure the rigor of this evaluation, we only use soft inflated objects of simple geometry here, i.e., balls, and interact with them through vertical point indentation. Besides the Pezzi ball and the yoga ball introduced before, we further include a beach ball with an anti-leakage valve to demonstrate our method's generality. Following Procedure~\ref{procedure2} first and then Procedure~\ref{procedure3}, their Young's moduli $E$ are estimated to be $2.34$ MPa, $1.34$ MPa and $4.10$ MPa, respectively. Note that the manometer is not used in this experiment, and the gauge pressures of all three balls are estimated via Procedure~\ref{procedure2} with previously calibrated $k_{s}=0.64$ in Section~\ref{sec:pressure}. As shown in Figure~\ref{fig:elasticity}b, we set up a z-axis sliding platform equipped with a mechanical force gauge to quantify the applied force and vertical displacement. The quantities that we compare, i.e., the height of the deformed ball $H$, the diameter of the upper circle $D_{u}$, the diameter of the circular contact region $D_{l}$, and the depth of the sunken region $d$, are selected for the following reasons: 1) they are representative of the deformed shape; 2) they exhibit large variations as the force increases; 3) they are easy to accurately measure. As summarized in Figure~\ref{fig:elasticity}c, the deformations of digitized balls simulated with estimated $P_{g}$ and $E$, in general, align well with the physical ones. The mean relative errors of the height, upper diameter, lower diameter, and depth of digitized balls, as compared to the real ones, are $6.2\%$, $2.1\%$, $10.1\%$ and $7.2\%$, respectively. Notably, the highest error $10.1\%$ lies in the lower diameter $D_{l}$, with the value from the simulation being slightly larger than the actual measurement. We argue that this discrepancy is mainly due to the fact that we do not model the friction between the object surface and the supporting plane. As a result, the flat contact region of the object is not restricted from extending horizontally and thus stretches more in simulation than in reality.

%%%%%%%%%%%%%%%%%%%%%%%%%%%%%%%%%%%%%%%%%%%%%%%%%%

\begin{figure*}[t]
    \centering
    \includegraphics[width=0.99\linewidth]{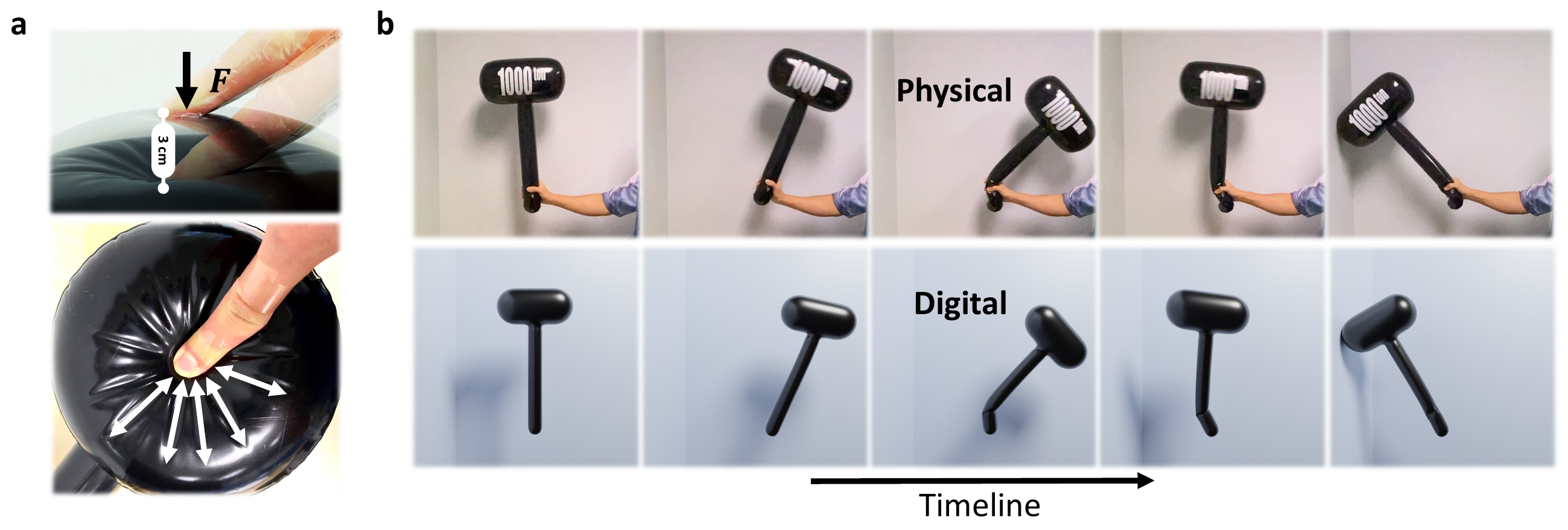}
    \caption{Extension to digitize an irregularly shaped inflated hammer toy. (a) Collection of $\{F, w(0)\}_{i=1}^{N_{1}}$ value pairs via finger-induced point indentation and observation of radial wrinkles. (b) The comparison between swinging the physical toy and its digital counterpart demonstrates spatio-temporally aligned motions and deformations.}
    \label{fig:hammer}
\end{figure*}

\subsection{Dynamic Behaviors of Digitized Objects}

Besides matching the quasi-static behaviors of real objects subject to external loading, it is also essential to replicate their deformations and motions in dynamic scenarios. A representative example in this respect is when they collide with other objects in the environment. Such interactions not only depend on the objects' physical properties but also involve complex energy transformation. The energy dissipation due to collision takes various forms (heat, sound, etc) and is still under active investigation in impact mechanics, which is out of the scope of this paper. Instead of explicitly modeling the energy loss, we use the coefficient of restitution (COR), the ratio between outgoing and incoming velocities, as a surrogate to characterize the energy variations. This adaptation is valid for inflated objects since they normally recover their original shapes and gauge pressures soon after the collision. For objects at low velocities, their COR is often assumed to be velocity-independent and approximated by a constant. Neglecting the air resistance~\footnote{We use a Pezzi ball of radius 0.13m and generate low-velocity motions (below 3m/s), air resistance ($\leq$ 0.13N) is negligible compared to the ball's gravity ($\sim$1.5N).}, we measure the COR of a soft inflated object through free fall $\text{COR} = \sqrt{h / H}$, where $h$ is the bounce height and $H$ is the drop height.

We first study the collision between balls and the ground through a free-fall experiment. We use the 240 frames per second (FPS) slow motion on iPhone to record the free fall of a Pezzi ball and compare its maximally deformed shape when it hits the ground to its digital replica. The free-fall simulation for the digitized Pezzi ball is created in the same way as in Section~\ref{sec:modulus}, with estimated $P_{g}$ and $E$. The gravity is applied based on the measured weight and computed volume of the real Pezzi ball. In addition, velocity damping is added when the digitized ball collides with the ground to account for the energy dissipation, and the damping strength is adapted to the measured COR. Figure~\ref{fig:bouncing}a shows that the digital replica nicely recovers the maximally deformed shape of the real Pezzi ball. Larger deformation when dropping from a higher position is also reflected. Using estimated $P_{g}$ and $E$ coupled with measured COR, we further reproduce a more general bouncing movement with the digitized Pezzi ball. The ball is launched with a horizontal velocity $v_{i}=3$ m/s and collides with the wall and the ground in sequence. As observed in Figure~\ref{fig:bouncing}b, the digitized ball faithfully replicates the overall trajectory and deforms accurately during collisions. A side-by-side temporal comparison of the bouncing experiments can be found in the accompanying video.

%%%%%%%%%%%%%%%%%%%%%%%%%%%%%%%%%%%%%%%%%%%%%%%%%%

\subsection{Generalization to Objects of Irregular Shape}

To demonstrate the generality of our method regardless of a uniform shape, we further apply our method to digitize an inflated hammer toy. Following Procedures~\ref{procedure2} and~\ref{procedure3} with previously calibrated $k_{s}=0.64$ in Section~\ref{sec:pressure}, its gauge pressure $P_{g}$ and surface elastic modulus $E$ are estimated to be 3.955 kPa and 53.98 MPa. Figure~\ref{fig:hammer}a illustrates how we poke the hammer toy via fingers (with haptic sensors mounted) to obtain $\{F, w(0)\}_{i=1}^{N_{1}}$ pairs.

During the experiments, we used the hammer toy to perform a simple smashing motion and replicate the same motion with its digital copy for evaluation. The motion begins with the hammer being held in a vertical position. We then rotate the hammer clockwise for 30 degrees around its base to get more space for a strong smashing. Next, we reverse the rotation direction and do another 60 degrees to hit the hammerhead onto the wall. The hammer ends with a 30-degree angle from the initial position. Regarding the simulation, we transfer the same rotation to the lower part of the handle. As visualized by the keyframes in Figure~\ref{fig:hammer}b, consistent motions and deformations between the physical hammer toy and its digital replica are observed. In particular, the digitized hammer exhibits realistic and accurate bending effects on both the hammer handle and the head-handle connection region. Note that the surface material of the hammer toy is much stiffer than that of the balls used before, and this is well reflected by the estimated Young's modulus $E=53.98$ MPa. A side-by-side temporal comparison of the hammer toy smashing motion can be found in the accompanying video.
\section{Limitations and Future Work}

As a first step toward pervasively digitizing our
physical surroundings while bypassing high-cost, time-consuming, and intrusive laboratory-based setups, this research admits the following limitations for future explorations.

The analytical inverse model formulated in Section \ref{sec:framework} is grounded in the correlation between the external force loading and the wrinkling patterns exhibited by the target deformable object. While the choice of localized modeling makes the proposed approach potentially generalizable to any deformable object with a small locally convex surface region, it is possible that the wrinkling patterns become less observable when the assumption of $W(0) \ll -1$ (the dimensionless indentation depth, please refer to Section \ref{sec:theory}) no longer holds or the linear scaling factor $k_{s}$ (please refer to Section \ref{sec:cap}) deviates from the regime of non-trivial point indentation. This could happen when the target object shows intricate geometry or has a very thick surface compared to its scale. In addition, the proposed framework is validated using objects of simple geometry and uniform materials only. Deformable objects showing more complex shapes and material compositions may exhibit locally variant deformation behaviors. Lastly, complex multi-object interaction (e.g., collision) requires additional energy-related considerations such as the coefficient of friction. 
Combining our analytical inverse model with dynamics-based friction measurements \cite{harnoy2008modeling} may shed light on generalizing to large-scale inter-object modeling.

\section{Conclusion}
\label{sec:discussion}

In this research, we address a long-standing problem in 3D digitization and interaction: how to non-invasively infer a deformable object's physical properties, complementary to the vision-based geometry and appearance reconstruction, to create its physically accurate digital replica?
The proposed framework computationally estimates two physical properties essential to digitizing deformable objects, gauge pressure and elastic modulus, by leveraging the correlation between force-induced indentation and resulting deformation patterns on the objects. 
Our approach only requires consumer-level sensors and simple user interventions, bypassing high-cost equipment, restrictive use cases, invasive operations, and tedious procedures. 
We hope this research serves as a practical and convenient digitization tool as well as provides insights for researchers to explore new avenues to advance physical realism in digital data creation and natural human-computer interaction.

%%%%%%%%%%%%%%%%%%%%%%%%%%%%%%%%%%%%%%%%%%%%%%%%%%%%%%%%%%%%%

% \clearpage
\bibliographystyle{ACM-Reference-Format}
\bibliography{paper.bib}

\end{document}